\documentclass[draft]{agujournal}
\journalname{Geophysical Research Letters}

\begin{document}

%% ------------------------------------------------------------------------ %%
%  Title
%
% (A title should be specific, informative, and brief. Use
% abbreviations only if they are defined in the abstract. Titles that
% start with general keywords then specific terms are optimized in
% searches)
%
%% ------------------------------------------------------------------------ %%

% Example: \title{This is a test title}

\title{ An axisymmetric limit for the width of the Hadley cell on planets with large obliquity and long seasonality}

\authors{I. Guendelman\affil{1} and Y. Kaspi\affil{1}}

\affiliation{1}{Weizmann Institute of Science, Department of Earth and Planetary Sciences, Rehovot, Israel}

\correspondingauthor{Ilai Guendelman}{ilai.guendelman@weizmann.ac.il}

%% Keypoints, final entry on title page.

% Example:
% \begin{keypoints}
% \item	List up to three key points (at least one is required)
% \item	Key Points summarize the main points and conclusions of the article
% \item	Each must be 100 characters or less with no special characters or punctuation
% \end{keypoints}

%  List up to three key points (at least one is required)
%  Key Points summarize the main points and conclusions of the article
%  Each must be 100 characters or less with no special characters or punctuation

\begin{keypoints}
\item = The Hadley cell ascending branch position in planets with strong seasonal variation of temperature is mainly bounded by the rotation rate.
\item = A similar rotation rate dependence arises in the axisymmetric theory.
\item = This theory can explain the ascending branch position on Mars and Titan.
\end{keypoints}

%% \begin{abstract} starts the second page

\begin{abstract}
Hadley cells dominate the meridional circulation of terrestrial atmospheres. The Solar System terrestrial atmospheres, Venus, Earth, Mars and Titan, exhibit a large variety in the strength, width and seasonality of their Hadley circulation. Despite the Hadley cell being thermally driven, in all planets, the ascending branch does not coincide with the warmest latitude, even in cases with very long seasonality (e.g., Titan) or very small thermal inertia (e.g., Mars). In order to understand the characteristics of the Hadley circulation in case of extreme planetary characteristics, we show both theoretically, using axisymmetric theory, and numerically, using a set of idealized GCM simulations, that the thermal Rossby number dictates the character of the circulation. Given the possible variation of thermal Rossby number parameters, the rotation rate is found to be the most critical factor controlling the circulation characteristics. The results also explain the location of the ascending branch on Mars and Titan.
\end{abstract}

\section{Introduction}
\label{sec:intro}

Observations and models show that the Hadley circulation varies considerably between the Solar System terrestrial atmospheres of Venus, Earth, Mars and Titan. Venus' lower atmosphere is composed of two hemispherically symmetric equator-to-pole Hadley cells \citep{read2013dynamics, sanchez2017atmospheric}. On Earth, similar to Venus, Hadley cells exist in both hemispheres, but due to Earth's obliquity, the ascending branch latitude and the strength of the two cells vary seasonally, where during the solstice there is a strong and wide winter cell and a narrow and weak summer cell \citep[e.g.,][]{dima2003seasonality}. 

Both Mars and Titan, exhibit stronger seasonality in the Hadley circulation compared to Earth, despite the fact that the obliquity of Mars and Titan is similar to Earth's obliquity. The strong seasonality on Mars is due to its thin atmosphere and rocky surface resulting in a low thermal inertia and a short radiative timescale. Mars' Hadley circulation transits from two hemispherically symmetric cells at equinox, to one solstice cell, with air rising at midlatitudes \citep{read2015physics}. Thus, although at solstice, Mars' maximum surface temperature is at the pole, the Hadley cell ascending branch does not reach the pole. 

Titan's tropospheric radiative timescale is considerably longer than its orbital period \citep{mitchell2016climate}, which explains why Titan's maximum surface temperature seem to stay near the equator during the seasonal cycle \citep{jennings2016temp, lora2015titan}. However, observations of Titan's methane clouds, show a significant seasonal cycle as they shift from one pole to the other during Titan's year \citep{brown2002direct, turtle2011seasonal, roe2012titan, trutle2018titan}. Different models associate the polar clouds to different phenomena. \citet{schneider2012titan} associate the polar clouds to the meridional convergence (analogous to the inter tropical convergence zone, ITCZ, on Earth) indicating a pole-to-pole Hadley circulation \citep[e.g,][]{roe2012titan}. In contrast, \citet{lora2015titan} and other studies \citep[e.g.,][]{mitchell2006dynamics, mitchell2009titan, mitchell2008drying, newman2016titan} relate the polar clouds to intensive polar warming during solstice, while the meridional convergence occurs at midlatitudes. 

The variability of the terrestrial atmospheric circulation within the Solar System, is a result of the variability in the planets' orbit, rotation rate, atmospheric mass, radius etc. Different studies explored the effect of different planetary parameters on the atmospheric circulation, showing that the large scale circulation depends greatly on the planetary parameters and atmospheric characteristics \citep[e.g., ][]{ferreira2014climate, kaspi2015atmospheric, linsenmeier2015climate, faulk2017effects, chemke2017dynamics}.

More specifically, \citet{faulk2017effects} studied the dependence of the meridional circulation seasonal cycle on the rotation rate using an idealized aquaplanet GCM, showing that in the case of an Earth-like rotation rate, the ITCZ and the ascending branch of the Hadley circulation do not reach the pole, even in an eternal solstice case, where the maximum temperature is at the pole. This result, together with Mars' ascending branch not reaching the pole, even though its maximal surface temperature is at the pole \citep{forget1999improved, read2015physics} and some Titan models predicting the ascending branch being poleward from the warmest latitude \citep[e.g.,][]{lora2015titan}, is puzzling. Theoretical expectations are that the Hadley cell ascending branch, being a thermally driven circulation, will follow the warmest latitude \citep{neelin1987modeling}, or the latitude of maximum low level moist static energy \citep{emanuel1994large, prive2007monsoon}, which is not the case for Mars and neither for the \citet{faulk2017effects} simulations. This study shows that axisymmetric theory \citep{held1980nonlinear, lindzen1988hadley, caballero2008axisymmetric} has a similar rotation rate dependence as the modeling results and the observations on Mars and Titan show. 

For Earth, there are several other theories regarding the Hadley circulation and the ITCZ position, that unlike the axisymmetric theory take into account the eddy contribution, and involve processes such as the flux of eddies across the equator \citep[e.g.,][]{kang2008response,bischoff2014energetic, adam2016itcz1, adam2016itcz2, wei2018energetic}, moist processes \citep[e.g.,][]{neelin1987modeling}, baroclinicity \citep{held2000} and supercriticality \citep{korty2008extent,levine2015baroclinic}. However, in this study which aims to understand the leading order effects over a wide range of conditions, we focus on the simpler, axisymmetric theory, as a leading order theory for the zonally symmetric climate balance. In section \ref{sec:theory} we derive the axisymmetric theory for the solstice case following \citet{lindzen1988hadley} (hereafter LH88), and solve it numerically to include a wide range of planetary parameters. In section \ref{sec:model} we briefly describe the numerical model and present the simulation results, relating them to the axisymmetric theory. In section \ref{sec:conc} we discuss the results and their implication for the Solar System atmospheres. 

\section{Axisymmetric theory}
\label{sec:theory}

The axisymmetric theory introduced by \citet{held1980nonlinear} and further developed by LH88 to include the solstice case, is a theory for the Hadley circulation that neglects eddy contribution and diffusive processes. Despite the importance of eddies \citep[e.g.,][]{walker2006eddy}, the axisymmetric theory has been found to overall give a good leading order estimate to the cell extent. Following LH88, angular momentum conservation at the top of the cell is assumed, and the angular momentum conserving wind, at latitude $\phi$, of an air parcel starting at rest from latitude $\phi_1$ (the ascending branch of the Hadley cell) is
\begin{linenomath*}
\begin{equation}
\label{eq:uM}
u_M=\Omega a \frac{\cos^2\phi_1-\cos^2\phi}{\cos\phi},
\end{equation}
\end{linenomath*}
where $\Omega$ is the planetary rotation rate and $a$ is the planetary radius. Assuming that the flow is in cyclostrophic balance and that thermal wind balance holds to leading order, Eq. \ref{eq:uM} together with hydrostatic balance, results in an expression for the angular momentum conserving potential temperature ($\theta$) 
\begin{linenomath*}
\begin{equation}
\label{eq:thetam}
\frac{\theta(\phi)-\theta(\phi_1)}{\theta_0}=-\frac{\Omega^2a^2}{2gH}\frac{(\sin^2\phi-\sin^2\phi_1)^2}{\cos^2\phi},
\end{equation}
\end{linenomath*}
where $\theta_0$ is some reference potential temperature and $H$ is the troposphere height. Equation \ref{eq:thetam} can be expressed using the thermal Rossby number $R_t=\frac{2gH\Delta_H}{\Omega^2a^2}$ \citep{held1980nonlinear} to give
\begin{linenomath*}
\begin{equation}
\label{eq:thetam_rt}
\frac{\theta(\phi)-\theta(\phi_1)}{\theta_0}=-\frac{\Delta_H}{R_t}\frac{(\sin^2\phi-\sin^2\phi_1)^2}{\cos^2\phi},
\end{equation}
\end{linenomath*}
where $\Delta_H$ is the meridional fractional change of the radiative equilibrium temperature (LH88 and Eq. \ref{eq:thetae}). Taking a small angle approximation, the width of the circulation in the equinox case is $\propto R_t^{1/2}$ \citep{held1980nonlinear} and for the solstice case $\propto R_t^{1/3}$ \citep{caballero2008axisymmetric}.

In order to find the Hadley circulation edges, namely, the latitudes of the ascending and descending branches, we assume that the cells are energetically closed, that the temperature at the edge of the cells is continuous and that outside of the Hadley circulation the temperature is a radiative equilibrium temperature $\theta_e$
\begin{linenomath*}
\begin{equation}
\label{eq:thetae}
\frac{\theta_e}{\theta_0}=1+\frac{\Delta_H}{3}(1-3(\sin\phi-\sin\phi_0)^2),
\end{equation}
\end{linenomath*}
where $\phi_0$ is the latitude of maximum $\theta_e$. The energetically closed cell and temperature continuity assumptions translate to the following set of equations
\begin{linenomath*}
\begin{eqnarray}
\label{eqset1}
\int_{\phi_w}^{\phi_1} (\theta-\theta_e)\cos\phi d\phi &=& 0, \\
\int_{\phi_1}^{\phi_s} (\theta-\theta_e)\cos\phi d\phi &=& 0, \\
\theta(\phi_1)&=&\theta_e(\phi_1),\\
\theta(\phi_w)&=&\theta_e(\phi_w),\\
\label{eqset2}
\theta(\phi_s)&=&\theta_e(\phi_s),
\end{eqnarray} 
\end{linenomath*}
where $\phi_s$ and $\phi_w$ are the latitudes of the Hadley cell descending branch in the summer and winter hemispheres (edges of the circulation), respectively.
\begin{figure}[ht!]
\centerline{\includegraphics[height=10cm]{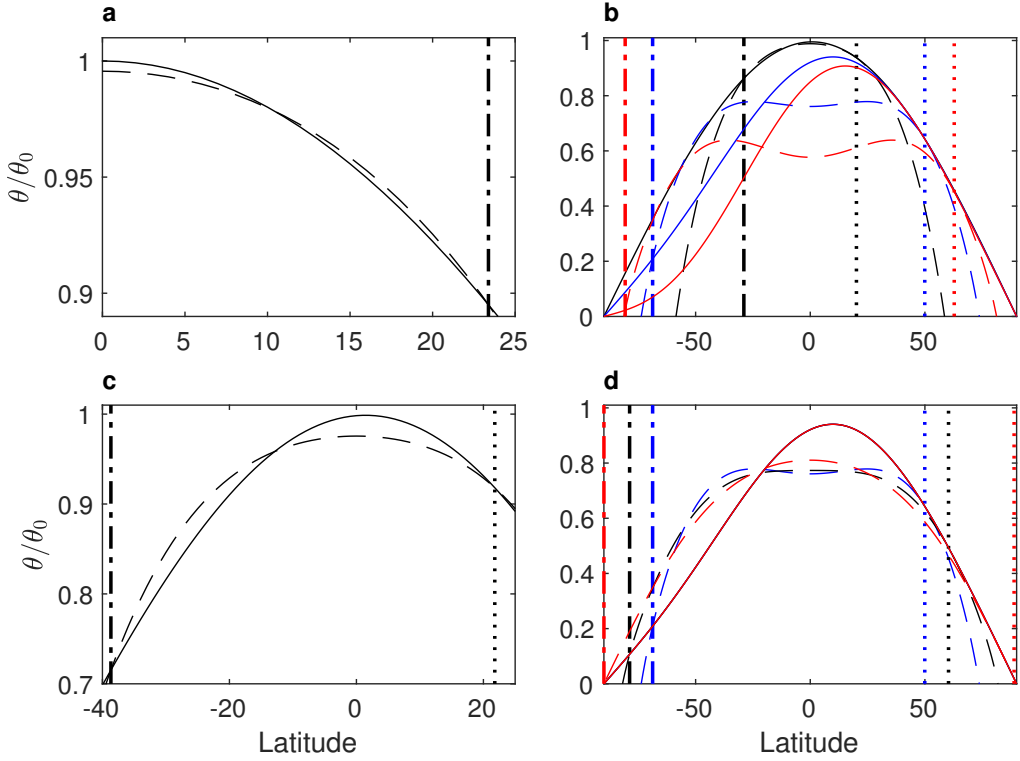}}
\caption{The radiative equilibrium potential temperature and the angular momentum temperature both multiplied by cosine of latitude ($\theta_e\cos\phi/\theta_0$, solid, and $\theta\cos\phi/\theta_0$, dashed, respectively), and normalized by maximum $\theta_e/\theta_0$. The vertical dotted and dashed dotted lines are for the winter cell ascending and descending branches, respectively. Panels a and c are similar to Fig. 5 in LH88, where panel a is the solution for the hemispherically symmetric case ($\phi_0=0^{\circ}$), and panel c is the solution for $\phi_0=6^{\circ}$ with $\Delta_H=1/6$ and Earth-like rotation rate. In panels b and d, $\phi_0=45^{\circ}$, where in panel b each color represents a different value of $\Delta_H$ with the values $0.01$ (black), $1/6$ (blue) and $1/3$ (red), all with Earth-like rotation rate. Panel d shows the solutions for different rotation rates, $0.5$ (red) $0.75$ (black) and $1$ (blue) all in Earth rotation rate units, with $\Delta_H=1/6$.}
\label{figtwo}
\end{figure} 
The unknowns that equations (\ref{eqset1})-(\ref{eqset2}) solve for are the ascending and descending branches latitudes $\phi_1$, $\phi_s$, $\phi_w$ and the temperature at the ascending branch $\theta(\phi_1)$. Graphically, the energetically closed cell assumption translates to an equal area between the angular momentum conserving and the radiative equilibrium temperature curves inside each cell. Figure \ref{figtwo} depicts the angular momentum conserving (dashed) and the radiative equilibrium (solid) temperature curves, multiplied by $\cos\phi$ for different cases, depicting the closed cell argument. Figures \ref{figtwo}a and \ref{figtwo}c are similar to Figure 5 in LH88 with the difference that here $\theta$ and $\theta_e$ are multiplied by $\cos\phi$, as the small angle approximation is not appropriate in this case. Fig.~\ref{figtwo}a shows the hemispherically symmetric cell and Fig.~\ref{figtwo}c shows the $\phi_0=6^{\circ}$ case, representing an Earth-like scenario. Figures \ref{figtwo}b and \ref{figtwo}d are for different temperature gradients and different rotation rates, respectively, where the latitude of maximum radiative temperature is at latitude $45^{\circ}$. All plots in Fig.~\ref{figtwo} show the position of the winter cell ascending (dotted line) and descending (dashed-dotted line) branches. Only the winter cell is shown, as for strong seasonal cases, which are the focus of this study, a summer cell barely exists. Comparing between Figures \ref{figtwo}b and \ref{figtwo}d, shows that slowing down the rotation rate is more efficient in widening the circulation than increasing the temperature gradient.  

\begin{figure}[ht!]
\centerline{\includegraphics[height=7cm]{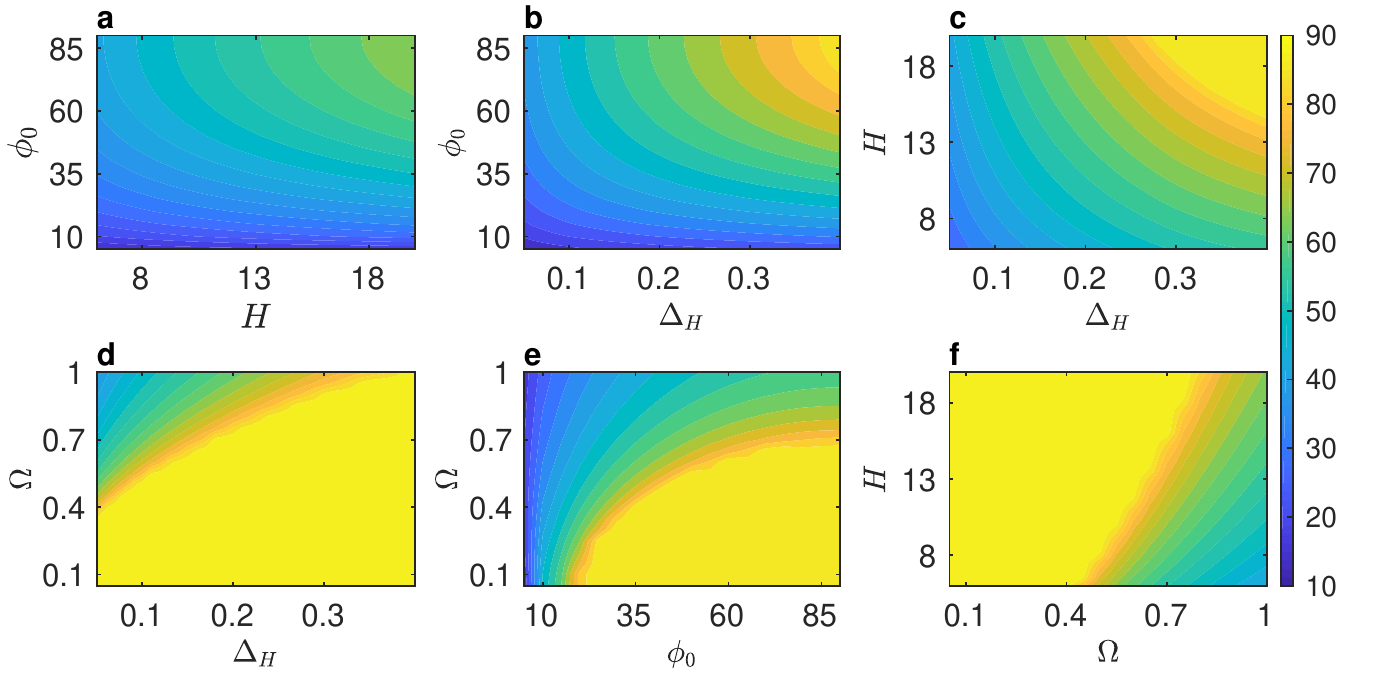}}
\caption{Axisymmetric solutions for the ascending branch latitude (color), in the parameter space of $\phi_0,\Delta_H,H$, and $\Omega$. The parameters are $\phi_0$, the latitude of maximum radiative equilibrium temperature, ranging from $5^{\circ}$ to $90^{\circ}$, with a value of $90^{\circ}$ when kept constant. The meridional radiative equilibrium temperature gradient, $\Delta_H$, ranging from $0.01$ to $0.4$, with a value of $0.17$ when kept constant. The tropopause height, $H$, ranging from $6$ to $20$ km, with a value of $15$ km when kept constant. The rotation rate, $\Omega$, raging from $0.05$ to $1$ in Earth's rotation rate units, with a value of $1$ when kept constant.}
\label{figthree}
\end{figure} 

The axisymmetric theory solutions shown in Figure \ref{figtwo} together with equations (\ref{eq:thetam}) and (\ref{eq:thetae}), show that the latitudes of the ascending and descending branches depend on different parameters. Solving numerically equations (\ref{eqset1})-(\ref{eqset2}) for a wide range of $\Omega,\Delta_H,\phi_0$ and $H$ values (Fig. \ref{figthree}) shows a clear difference between cases where the rotation rate is slowed down (Fig.~\ref{figthree}d-f), where the ascending branch easily reaches the pole, and cases where the rotation rate is kept with an Earth-like value (Fig.~\ref{figthree}a-c). This demonstrates that an Earth-like rotation rate or faster, limits the expansion of the circulation, such that even if $\phi_0$ is at the pole, and, for example, $\Delta_H$ is increased (over a realistic range), it is unlikely for $\phi_1$ to reach the pole unless the rotation rate is slowed down. This result is consistent with the simulations of \citet{faulk2017effects}. The choice of parameter values is guided by the observed values in the Solar System. $\Delta_H$, the normalized horizontal temperature difference, gets its largest value for Mars \citep[$\sim 0.4$, e.g.,][]{read2015physics}, and lowest value for Venus \citep[with nearly zero temperature gradient, e.g.,][]{read2013dynamics}. The tropopause height, $H$, taken to be be the circulation height scale \citep{walker2006eddy}, is highest on Titan and Mars reaching to $\sim20$ km \cite[e.g.][]{read2015physics, lora2015titan}.

\begin{figure}[htb!]
\centerline{\includegraphics[height=10cm]{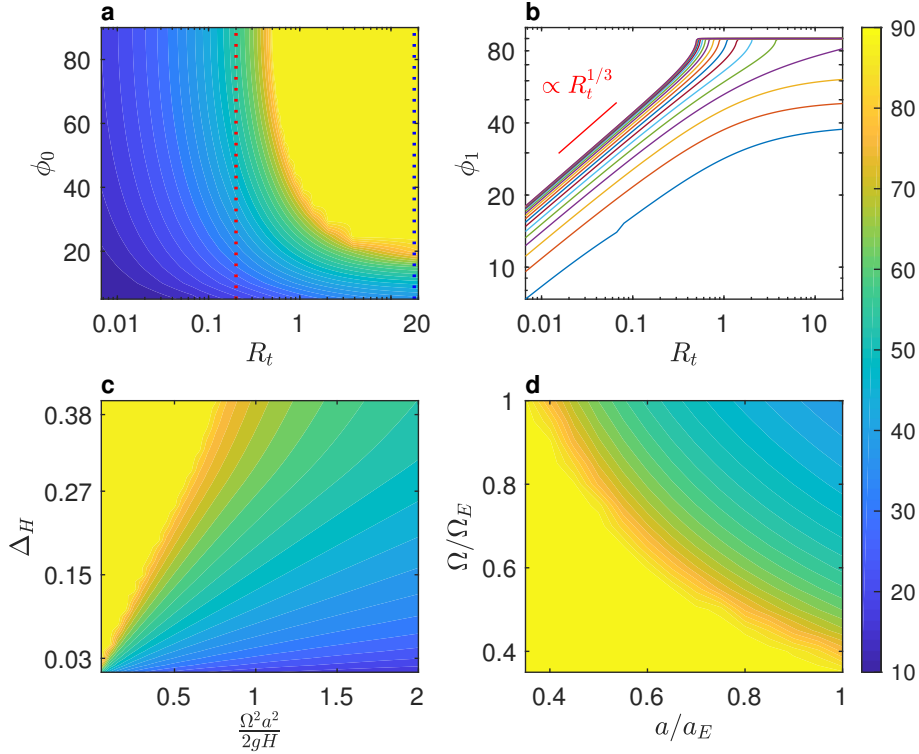}}
\caption{The latitude of ascending branch as a function of $\phi_0$, the thermal Rossby number, $R_t$, and its decomposition. (a) The ascending branch as a function of $\phi_0$ and $R_t$, where the dotted blue and red lines represent the $R_t$ value of Titan and Mars, respectively. Each curve in (b) is for a different $\phi_0$ value raging from $90{^\circ}$ (top) to $5^{\circ}$ (bottom) in a log-log plot of $\phi_1$ as a function of $R_t$, showing the correspondence to the \citet{caballero2008axisymmetric} scaling. (c) and (d) are the ascending branch latitude (color) as a function of $\frac{\Omega^2a^2}{2gH}$ and $\Delta_H$ and for different values of $a$ and $\Omega$, respectively, where $a_E$ and $\Omega_E$ are Earth's radius and rotation rate. In (c) $g=3$ ms$^{-2}$ is used, which is between the Mars and Titan surface gravity values.}
\label{figsix}
\end{figure} 

In order to understand this rotation rate dependence, we plot the ascending branch latitude as a function of $R_t$ and $\phi_0$ (Fig. \ref{figsix}a), showing that for each value of $\phi_0$ the position of the ascending branch is strongly dependent on $R_t$ (Fig.~\ref{figsix}a,b). For small values of $R_t$ there is a good agreement with the  \citet{caballero2008axisymmetric} scaling (Fig~\ref{figsix}b). The values of $R_t$ for the Solar System terrestrial atmospheres vary from $0.06$ on Earth to $370$ on Venus, with the value for Mars being $0.2$ and for Titan $18$ \citep{read2011dynamics}. Decomposing $R_t$ into $\frac{\Omega^2a^2}{2gH}$ and $\Delta_H$, which is a natural decomposition to a dynamical component ($\frac{\Omega^2a^2}{2gH}$) and a radiative one ($\Delta_H$), shows the range of possible values of $\frac{\Omega^2a^2}{2gH}$ is larger compared to that of $\Delta_H$ (Fig~\ref{figsix}c). As a result, this factor will have a larger role in limiting the width of the circulation. Examining the elements in $\frac{\Omega^2a^2}{2gH}$ shows a strong dependence on the rotation rate and radius (Fig. \ref{figsix}d). Taking the Solar System terrestrial atmospheres as a proxy, and comparing between the range of the different parameters in $R_t$, shows that the rotation rate is the only parameter known to vary by two orders of magnitude (Earth and Venus), while all other parameters vary by one order of magnitude or less. This together with the strong dependence of $R_t$ on the rotation rate is what makes the rotation rate the limiting factor on the circulation extent. Also, taking a closer look at the radius dependence, shows that it is not as strong as the rotation rate dependence, considering the surface gravity dependence on the planetary radius $g=4\pi G\rho a/3$. Here $\rho$ is the planet's mean density, which is a more fundamental characteristic of the planet than its surface gravity, and $G$ is the universal gravitational constant. Therefore, a more useful form to write the thermal Rossby number in this context, is 
\begin{linenomath*}
\begin{equation}
\label{eq:troalter}
R_t= \frac{8\pi\rho G H\Delta_H}{3\Omega^2a}. 
\end{equation}
\end{linenomath*}
Expressing $R_t$ in this form emphasizes the circulation dependence on the rotation rate.
\section{Idealized GCM simulations}\label{sec:model}
\subsection{Model description}
In order to test the theoretical framework presented in section \ref{sec:theory} in a more complete model, we use an idealized moist aquaplanet GCM \citep{frierson2006gray}, based on the GFDL dynamical core \citep{gfdl2004new}, used in this context by \citet{faulk2017effects}. The model radiation scheme is augmented to include a diurnal mean seasonal insolation dependence on obliquity \citep{pierrehumbert2010principles}. Similar to \citet{faulk2017effects} the atmospheric optical depth is constant with latitude. The model solves the primitive equations with a horizontal spectral grid of $2.8^{\circ}\times 2.8^{\circ}$ (T42) and $25$ uneven vertical levels. To analyze the climate we use a $50$ year climatology after reaching a statistical steady state. Figure \ref{figfour}a shows the model results for Earth-like parameters. The model shows a generally similar climate as Earth's \citep{kaspi2015atmospheric}.

\subsection{Simulation results}

\begin{figure}[htb!]
\centerline{\includegraphics[height=7cm]{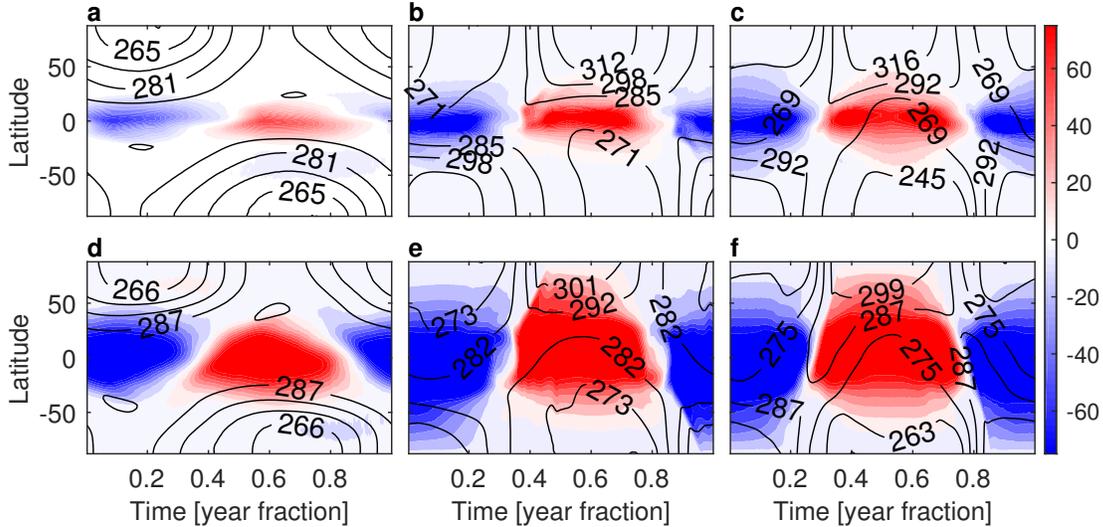}}
\caption{Hovmoller diagram of the streamfunction, $\psi=2\pi a\int vdp/g$, where $v$ is the zonal mean merdional wind, at its height of global maximum value (color, units $10^{10}$ kg s$^{-1}$) and surface temperature (contours, K) for different climates. The abscissa is a time axis, showing the year fraction. Top row simulations are with an Earth-like rotation rate and the bottom row simulations are with 1/4 of Earth's rotation rate. (a) and (d) are Earth-like simulations with obliquity $23^{\circ}$ and an Earth-like orbital period. (b) and (e) Simulations with obliquity $90^{\circ}$ and an Earth-like orbital period. (c) and (f) Simulations with obliquity $90^{\circ}$ and 4 times Earth's orbital period.}
\label{figfour}
\end{figure}
\begin{figure}[ht!]
\centerline{\includegraphics[height=7cm]{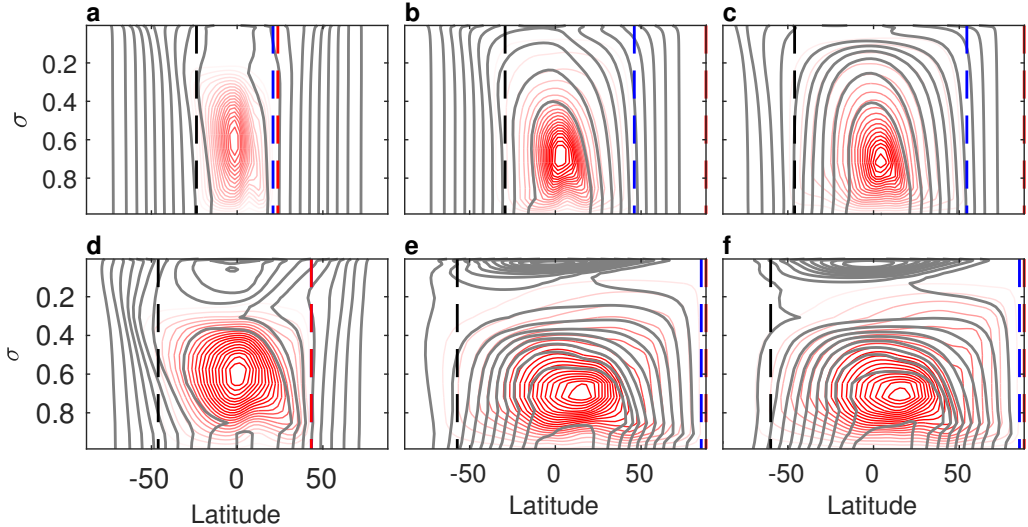}}
\caption{Streamfunction (red contours) and angular momentum (gray contours) vertical structure during southern hemisphere summer solstice, for different climates. Blue and Black vertical dashed lines are for the positions of the ascending and descending branches, respectively, defined to be the latitudes where the streamfunction reaches $5\%$ of its maximum \citep{walker2006eddy,faulk2017effects}. The red dashed vertical line is for the position of the maximum surface temperature. The simulation parameters for the different panels are the same as in Figure \ref{figfour}.}
\label{figfive}
\end{figure}

Three simulations with different degree of seasonality: Earth-like, obliquity $90^{\circ}$ with an Earth-like orbital period and obliquity $90^{\circ}$ with four times Earth's orbital period, are repeated with an Earth-like rotation rate and 1/4 of Earth's rotation rate. Figures \ref{figfour} and \ref{figfive} show that shifting the maximal temperature poleward from a reference Earth-like state (Fig.~ \ref{figfour}a), does not result in a global pole-to-pole Hadley circulation for simulations with an Earth-like rotation rate (Fig.~\ref{figfour}b), even when the temperature gradient is increased (Fig.~\ref{figfour}c). However, slowing down the rotation rate, allows the ascending branch to reach the pole, similar to \citet{faulk2017effects}. These results coincide with the theoretical solution of equations (\ref{eqset1})-(\ref{eqset2}) (Fig.~\ref{figthree}), where the ascending branch, for a realistic range of $\Delta_H$, does not reach the pole for an Earth-like rotation rate.

Figure \ref{figfive} shows that during the solstice of the strong seasonal cases, the meridional streamfunction follows the angular momentum contours (Fig.~\ref{figfive}b-f), implying that the eddy contribution is small. This means that despite the importance of eddies in the more Earth-like cases for the extent of the Hadley circulation \citep[e.g.,][]{walker2005response, walker2006eddy,korty2008extent} eddies seem to play less of a role in these cases. This alignment, relates to a previously suggested regime transition between an eddy mediated circulation at equinox, to a thermally driven one at solstice, where eddies do not contribute, suggesting that the use of axisymmetric theory is appropriate \citep{bordoni2008monsoons, bordoni2010regime, merlis2013hadley, geen2017regime}. Aside from the seasonal regime transition, there is a rotation rate related transition, where by slowing down the rotation rate, the streamfunction follows angular momentum contours. This regime transition in both rotation rate and seasonality is a result of weaker eddy momentum flux convergence that in turn allow the streamfunction to follow the angular momentum contours \citep{faulk2017effects}. Consistent with the angular momentum conserving cell these simulations do not exhibit superrotation. However, simulations with slower rotation rates (more similar to Titan and Venus) may exhibit superrotation \citep[e.g.,][]{kaspi2015atmospheric}, though the existence of superrotation together with a strong seasonal cycle is complex \citep{mitchell2014superrotation}.
\section{Discussion and conclusion}
\label{sec:conc}
Previous studies showed that as a planet rotates faster, the Hadley circulation contracts, the streamfunction becomes multicellular and the number of jets increases \citep{navarra2002numerical, walker2006eddy, kaspi2015atmospheric, chemke2015poleward, chemke2015latitudinal}. \citet{faulk2017effects}, studying the effect of the rotation rate in a seasonal cycle, showed that for a planet with an Earth-like rotation rate the Hadley cell ascending branch and the latitude of the ITCZ do not reach the pole, even when the maximum surface temperature is at the pole and the seasonal cycle is very long. 

Similar to \citet{faulk2017effects}, using an idealized GCM with different degrees of seasonality we show that for Earth-like rotation rate cases the Hadley cell ascending branch does not reach the pole (Figures \ref{figfour} and \ref{figfive}). A similar rotation rate limitation arises from the axisymmetric theory, predicting that the Hadley cell ascending branch latitude is limited for Earth-like rotation rate cases (Figure \ref{figthree}). This rotation rate dependence is a result of the angular momentum conservation and thermal wind assumptions that makes the width of the circulation to be a function of the thermal Rossby number (Figure \ref{figsix}). The quadratic dependence of the thermal Rossby number on the rotation rate (Eq.~\ref{eq:troalter}), and the limited range the other thermal Rossby number parameters exhibit in the Solar System planetary atmospheres, imply that the strongest limiting factor in controlling the ascending branch of the Hadley circulation is the rotation rate.

Studying these extreme cases, and the climate dependence on different planetary parameters, gives insight to the expected climate on other planetary atmospheres. Our Solar System terrestrial atmospheres are a good example for a variety of circulations, due to their large variability in planetary characteristics. Of particular interest is the seasonality on  Mars and Titan, both exhibiting a different circulation response to the seasonally varying surface temperature \citep{read2015physics, mitchell2016climate}. During the Martian solstice, maximum surface temperature is at the pole, however, the Hadley cell ascending branch is located at midlatitudes \citep{read2015physics}, consistent with the axisymmetric theory using Mars' $R_t$ (red dotted line in Fig.~\ref{figsix}a). 

On Titan, observational studies show that the maximum surface temperature stays around the equator during Titan's year \citep{jennings2016temp}; yet, cloud observations show a significant seasonal variation \citep[e.g.,][]{trutle2018titan}. Models of Titan's climate vary depending on their physical aspects \citep{horst2017titan}, with some models associating polar clouds with the Hadley cell ascending branch \citep[e.g.,][]{schneider2012titan} while others locate the ascending branch at midlatitudes \citep[e.g.,][]{lora2015titan}. The warmest latitude also varies between models \citep[e.g., the difference between dry and moist cases in][]{newman2016titan}. Particularly, \citet{lora2015titan} is an interesting case, where the peak surface temperature stays close to the equator while the ascending branch is located poleward, at midlatitudes. This variety of models can be explained using the axsymmetric theory. Following the blue dotted line in Fig.~\ref{figsix}a, which represents the $R_t$ value of Titan, we indeed find that if $\phi_0$ is taken to be $\sim 10^{\circ}$ the position of the ascending branch is at $\sim 45^{\circ}$,  in a general agreement with \citet{lora2015titan}. Also if $\phi_0\geq 30$ the ascending branch is predicted to be at the pole, similar to the dry case in \citet{newman2016titan}.

%  ACKNOWLEDGMENTS
%
% The acknowledgments must list:
%
% •	All funding sources related to this work from all authors
%
% •	Any real or perceived financial conflicts of interests for any
%	author
%
% •	Other affiliations for any author that may be perceived as
% 	having a conflict of interest with respect to the results of this
% 	paper.
%
% •	A statement that indicates to the reader where the data
% 	supporting the conclusions can be obtained (for example, in the
% 	references, tables, supporting information, and other databases).
%
% It is also the appropriate place to thank colleagues and other contributors.
% AGU does not normally allow dedications.

\acknowledgments
We thank Rei Chemke for fruitful conversations and help with the model configuration. We also thank the reviewers that helped improve this manuscript. Datasets used in this manuscript is available on https://doi.org/10.5281/zenodo.1442928. The authors acknowledge support from the Minerva Foundation with funding from the Federal German Ministry of Education and Research, and from the Weizmann institute Helen Kimmel Center for Planetary Science. 

\listofchanges
%%%

\end{document}